\documentclass[sigplan]{acmart}

\acmConference[ICPP]{The International Conference on Probabilistic Programming}{2018}{Boston, MA, USA}
\acmISBN{} 
\acmDOI{} 
\startPage{1}

\setcopyright{none}

\usepackage{booktabs}   
\usepackage{subcaption} 
\usepackage{listings}
                        
\usepackage{hyperref}

\begin{document}

\title{Inference Over Programs That Make Predictions}         


\author{Yura Perov}
\affiliation{
  \department{Babylon Health}
  \streetaddress{60 Sloane Avenue}
  \city{London}
  \postcode{SW3 3DD}
  \country{United Kingdom}                    
}
\email{yura.perov@babylonhealth.com}          

\begin{abstract}
This abstract extends on the previous work~\cite{perov2015applications,perov2016automatic} on program induction~\cite{muggletonapproaches} using probabilistic programming. It describes possible further steps to extend that work, such that, ultimately, automatic probabilistic program synthesis can generalise over any reasonable set of inputs and outputs, in particular in regard to text, image and video data.
\end{abstract}


\maketitle

\section{Introduction}

Probabilistic programming provides a natural framework for program induction, in which models can be automatically generated in the form of probabilistic programs given a specification. The specification can be expressed as input-output pairs, e.g. as ``23-year male has lower right abdominal pain (LRAP) and vomiting'' => ``Appendicitis''; ``34-year female has stiff neck, severe headache'' => ``Meningitis'', etc. We expect that given a set of such input-output examples, a program induction system will generate a model, which is capable of predicting an output for a new input. In the ideal scenario, the prediction will be a distribution over values rather than a specific deterministic value, as in ``23-year male has LRAP and vomiting'' can be mapped to both ``Appendicitis'' and ``Gastroenteritis'' with probabilities $p_1, p_2$.

In the previous work~\cite{perov2015applications,perov2016automatic} it was shown how to define an adaptor~\cite{johnson2007adaptor}, strongly-typed grammar as a probabilistic program, which can generate other probabilistic programs given a specification in the form of a set of observations:
\begin{lstlisting}
(assume model (run-grammar 'array 'word))
(noisy-observe (apply model inp1) outp1)
...
(noisy-observe (apply model inpN) outpN)
(predict inp0)
\end{lstlisting}

By performing inference over $model$-s, it was possible to infer simple probabilistic programs (specifically, samplers from one dimensional distributions, e.g. Bernoulli, Poisson, etc.) that can generalise over any available training data (input and output pairs) and predict new outputs given new inputs. The results were comparable to another state-of-the-art approach of program induction, genetic programming.

\section{Further Extensions}

To facilitate the research into program induction using probabilistic programming, the follow improvements into the methodology are suggested:

\subsection{Using non-parametric, hierarchical distributions for the grammar}

In the previous work, the adaptor, strongly-typed grammar was used with an enhancement of adding an option of drawing variables from local environments. For example, if we were looking for a model that takes two integers as arguments and outputs another integer, we would define an initial local environment scope with two variables $x_1, x_2$ and $N$-predefined constants (like $\pi, e^{1}$, etc.), and sample a candidate program from the grammar, i.e. as from \lstinline[breaklines=true]|(grammar (localscope (typed 'x1 'int) (typed 'x2 'int)) 'int)|. The $(grammar \ldots)$ method randomly generates an expression that produces an {\it integer} (or any other type):

\begin{itemize}
\item either it will an integer constant (hence $expr_{int} -> c$),
\item an integer variable (hence $expr_{int} -> x_i$),
\item a predefined method $f$, which returns an integer, with $M$ arguments (for each of which $grammar$ will be called recursively to construct the full expression $(f\ expr\-arg1\ \ldots\ expr\-argM)$),
\item a new extended local scope (via $(let \ldots)$) with one more variable of any supported type (including functions themselves), such that the returned expression of that $(let ...)$ is still of integer type (hence $expr_{int}\ ->\ (let\ (newsymb\ expr_{*})\ expr'_{int})$),
\item short-circuit $if$ such that $expr_{int}\ ->$ \newline $\ (if\ expr_{bool}\ expr_{int}\ expr_{int})$,
\item recursive call to the current function $expr_{int}\ ->\ (recur\ \ldots)$ assuming the type signature of the current function is integer.
\end{itemize}

To make the priors over programs more flexible, we suggest to use non-parametric~\cite{mansinghka2012structured,johnson2007adaptor}, hierarchical~\cite{liang2010learning} priors. That is, instead of adding a local environment with a variable of arbitrary type and randomly grammar-generated expression of that type, we suggest to instead draw expressions from a non-parametric distribution defined as a $mem$oized function with a base function being the grammar itself. The arguments of a call to that function might be another expression generated using the same grammar, which ensures that the ``keys'' of that $mem$oized function will be generated based on the program induction inputs. The hierarchical property of such a prior might be achieved by ensuring that the same $mem$oized function might be call in its body; by doing so, we allow this function to decide whether to return an expression or to make another call to itself with different arguments, hence going deeper in the hierarchy.

\subsection{Extending types supported, including higher-order typing}

Another improvement over the previous work can be achieved by extending the types which can be used by the grammar. This includes adding more types like dictionaries, lists, matrices, sets, queues, etc. Also, ideally we would like to support ``recursive'' type definitions such that the grammar not just produces the expressions to be evaluated, but is also capable of producing expressions that generate other expressions to be evaluated.

\subsection{Using discriminative model proposals}

To facilitate inference in a probabilistic program induction framework, we can use modern advances in probabilistic programming inference. In particular, we can use discriminative models, such as neural networks, to facilitate the inference~\cite{perov2015data,gu2015neural,le2016inference,douglas2017universal}.

\subsection{Incorporating the compressed knowledge in the form of embeddings}

The set of functions that can be used by the grammar also can be extended. Specifically, we believe one of the most interesting additions into that set might be the pre-trained embeddings. For example, we can incorporate $word2vec$~\cite{mikolov2013efficient} functions which would map a type ``word'' to a type ``float''. This should allow the program induction to benefit from the compressed knowledge which the $word2vec$ and similar embedding models represent.

\subsection{``Ultimate'' task for the induction}

In the previous work, the probabilistic program induction was performed over a simple one dimensional distribution.

We believe that the most effective and cheap way to provide as much training data as possible is to set a task of predicting 1-20 words given previous 20-500 words for an ``arbitrary piece of text''. These pieces might be extracted from any source, e.g. from Wikipedia, news web-sites, books, etc. The observational likelihood might be a Binominal distribution $Bernoulli(N, p)$, where $N$ is the number of words to predict for that particular input-output pair, and $p$ is the probability of ``success''. This approach follows the methods of noisy Approximate Bayesian Computation~\cite{marjoram2003markov}. Parameter $p$ also might be varied in the process of inference, hence we might be performing simulated annealing. We believe that with enough computational power and with rich enough priors, the inference will be able to find models that predicts reasonably well what the next word or list of few words should be.

Once a good model that can predict next words is well trained, this task can be extended to predicting: audio and video~\cite{perov2015data} sequences, image reconstruction~\cite{mansinghka2013approximate,kulkarni2015picture}, text-to-image and image-to-text tasks, as well as then ultimately performing actions in environments like OpenAI Gym.

\subsection{Distributing the computations}

The inference over such a gigantic set of input-output pairs will require a massive amount of computations which needs to be distributed. One approach to run the inference in parallel might be running multiple $N >> 1$ Markov chains (e.g. using Metropolis-Hastings algorithm) where each chain is given some subset of observations (i.e. it would be similar to {\it stochastic gradient descent} approach), as well as those chains ``lazily'' share the hidden states of the non-parametric, hierarchical, adaptor grammar. By ``lazy'' sharing we mean that the hidden states of the non-parametric components of the grammar are to be updated from time to time.

\subsection{Discussion over proposals}

While this abstract has focused on possible enhancements to improve priors over models as well as possible ways of setting the inference objective, it is also important to allow the proposal over a new probabilistic program candidate (e.g., as in $X \to X'$ in Metropolis-Hastings) to be flexible, ideally by sampling the proposal function from the grammar as well. In that case, it will be ``inference over inference'', i.e. nested inference~\cite{rainforth2018nesting} over the grammar and over the proposal. Another way of improving the process of inducing a new program candidate is the merging of two existing programs as in~\cite{hwang2011inducing}.

\subsection{Conclusion}

This short abstract extends the previous work by suggesting some enhancements to allow more practical probabilistic program induction.

Implementing a system which is capable of such complex probabilistic program induction will require a lot of resources, with the computational resource and its distribution being the most expensive one, presumably.

Another careful consideration should be made to the choice of three languages:
\begin{itemize}
\item the language in which the system is to be written,
\item the language in which the grammar is defined,
\item the language of the inferred probabilistic programs (models).
\end{itemize}

It might be beneficial if it is the same language altogether, such that the system can benefit from recursive use of the same grammar components (e.g. for doing inference over the inference proposal itself as suggested before). Also, ideally it is a ``popular'' language (or a language that can be easily transformed into such), such that all publicly available source code made be incorporated~\cite{maddison2014structured} into the priors. Examples of such language candidates are Church~\cite{goodman2012church}, Venture~\cite{mansinghka2014venture}, Anglican~\cite{wood2014new,tolpin2015probabilistic}, Probabilistic Scheme~\cite{paige2014compilation} or WebPPL~\cite{goodman2014design}.

\begin{acks}                            

This abstract is the extension to the work~\cite{perov2015applications}, as well as incorporates ideas which had been published online~\cite{perov2018website_one,perov2018website_two,perov2018website_three}.

\end{acks}

\bibliography{refs.bib}



\end{document}